\documentclass[conference]{IEEEtran}

\input{Qcircuit}
\usepackage{graphicx} 
\usepackage{amsmath}
\usepackage{amssymb}
\usepackage{amsthm}
\usepackage{subcaption}
\usepackage{float}
\usepackage{epstopdf}
\usepackage{multicol}
\usepackage{multirow}
\usepackage{booktabs}
\usepackage[autostyle=false, style=english]{csquotes}
\MakeOuterQuote{"}
\usepackage{color}
\usepackage{algorithm}

\usepackage{tikz}
\usetikzlibrary{decorations.pathreplacing,calc}

\ifCLASSINFOpdf
\else
\fi

\begin{document}
%
\title{Quantum Carry Lookahead Adders for NISQ and Quantum Image Processing}

\author{\IEEEauthorblockN{ Himanshu Thapliyal, Edgard Mu\~{n}oz-Coreas and Vladislav Khalus 
\IEEEauthorblockA{Department of Electrical and Computer Engineering\\
University of Kentucky, Lexington, KY, USA \\
Email: hthapliyal@ieee.org}
}

}

\bstctlcite{IEEEexample:BSTcontrol}

%


\maketitle

\begin{abstract}

Progress in quantum hardware design is progressing toward machines of sufficient size to begin realizing quantum algorithms in disciplines such as encryption and physics.  Quantum circuits for addition are crucial to realize many quantum algorithms on these machines.  Ideally, quantum circuits based on fault-tolerant gates and error-correcting codes should be used as they tolerant environmental noise.  However, current machines called Noisy Intermediate Scale Quantum (NISQ) machines cannot support the overhead associated with fault-tolerant design. In response, low depth circuits such as quantum carry lookahead adders (QCLA)s have caught the attention of researchers.  The risk for noise errors and decoherence increase as the number of gate layers (or depth) in the circuit increases.  This work presents an out-of-place QCLA based on Clifford+T gates.  The QCLAs optimized for T gate count and make use of a novel uncomputation gate to save T gates.  We base our QCLAs on Clifford+T gates because they can eventually be made fault-tolerant with error-correcting codes once quantum hardware that can support fault-tolerant designs becomes available.  We focus on T gate cost as the T gate is significantly more costly to make fault-tolerant than the other Clifford+T gates.  The proposed QCLAs are compared and shown to be superior to existing works in terms of T-count and therefore the total number of quantum gates.   Finally, we illustrate the application of the proposed QCLAs in quantum image processing by presenting quantum circuits for bilinear interpolation.    

\end{abstract}


%
\IEEEpeerreviewmaketitle

\section{Introduction}

Quantum computing offers significant speedups for algorithms for encryption, searching and scientific computations \cite{Caraiman2012motivation} \cite{Novo2017Hamiltonian}.  Arithmetic units such as adders are needed to implementing many of these quantum algorithms.  Thus, researchers proposed adders for quantum computers \cite{Thapliyal2013qcla} \cite{Draper2006qcla}.   

Existing quantum computers (Noisy Intermediate Scale Quantum (NISQ) machines) are plagued by noise errors \cite{Rigetti2020quantumbox} \cite{Devitt2013Tgatebuild}.  Computations can fail due to quantum operation errors and quantum coherence errors \cite{Rigetti2020quantumbox}.  
To reduce impact from coherence errors, the depth (number of gate layers) should be minimized.

Thus, low depth arithmetic circuits (such as quantum carry lookahead (QCLA) adders) have been proposed (see \cite{Cheng2002QCLAinplace} \cite{Thapliyal2013qcla} \cite{Draper2006qcla}).  Out-of-place QCLAs (both inputs restored and sum on ancillae) such as \cite{Babu2013QCLAoutofplace} \cite{Draper2006qcla} and \cite{Thapliyal2013qcla} and in-place QCLAs (one input restored and sum replaces other input) such as \cite{Draper2006qcla} \cite{Thapliyal2013qcla} and \cite{Cheng2002QCLAinplace} are proposed.  However, these works suffer from overhead in terms of T gates and qubits.  The fault-tolerant implementation cost of the T gate is higher compared to other quantum gates (such as Clifford gates) \cite{Devitt2013Tgatebuild} \cite{Gidney20184TgateToffolibuild}.  By using recent developments such as improved Toffoli gate implementations (see \cite{Gidney20184TgateToffolibuild}) we can design QCLAs with low T gate and/or qubit overhead.  

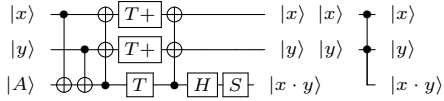
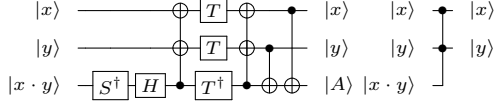
\begin{figure}
	\centering
	\begin{subfigure}[htbp]{3in}
		\scriptsize
		\[
		\Qcircuit @C=0.3em @R=0.5em @!R{
			\lstick{\ket{x}} & & \ctrl{2} &\qw &\targ & \gate{T+} & \targ & \qw & \qw & \qw & \qw &\qw &\rstick{\ket{x}} & & & & & & & & & & & & & & & \lstick{\ket{x}} & & \ctrl{1} & \qw & \rstick{\ket{x}}\\
			\lstick{\ket{y}} & & \qw &\ctrl{1} &\targ & \gate{T+} & \targ & \qw & \qw & \qw & \qw &\qw &\rstick{\ket{y}}  & & & & & & & &   & & & & & & & \lstick{\ket{y}} & & \ctrl{1} & \qw & \rstick{\ket{y}} \\
			\lstick{\ket{A}} & & \targ &\targ &\ctrl{-2} & \gate{T} & \ctrl{-2} & \gate{H} & \gate{S} &\qw &\rstick{\ket{x \cdot y}} & & & & & & & & & & & & &  & & & & & & \qwx &\qw &\rstick{\ket{x \cdot y}} \\  
		}
		\]
		\caption{The temporary logical-AND gate and its Clifford+T gate implementation.   T-count is $4$. The T gate needed to produce the $|A⟩$ state is included in the T gate cost of the logical AND gate (for a total of 4 T gates). Source: \cite{Gidney20184TgateToffolibuild}}
		\label{bi-logical-AND}
	\end{subfigure} \qquad  \begin{subfigure}[hb]{3in}
		\scriptsize
		\[
		\Qcircuit @C=0.4em @R=0.5em @!R{
			\lstick{\ket{x}} & & \qw & \qw & \qw & \targ & \gate{T} & \targ & \qw & \ctrl{2} & \qw & \rstick{\ket{x}} & & & & & & & & & & & & & & & \lstick{\ket{x}} & & \ctrl{1} & \qw & \rstick{\ket{x}}\\
			\lstick{\ket{y}} & & \qw & \qw & \qw & \targ & \gate{T} & \targ & \ctrl{1} & \qw &  \qw & \rstick{\ket{y}} & & & & & & & & & & & & & & & \lstick{\ket{y}} & & \ctrl{1} & \qw & \rstick{\ket{y}} \\
			\lstick{\ket{x \cdot y}} & & \qw &  \gate{S^{\dag}} &  \gate{H} & \ctrl{-2} & \gate{T^{\dag}} & \ctrl{-2} & \targ & \targ &\qw  &\rstick{\ket{A}} & & & & & & & & & & & & & & & \lstick{\ket{x \cdot y}} &  & \qw & & & \\
		}
		\]
		\caption{The proposed uncomputation gate and its Clifford+T gate implementation.   T-count is $3$.  }
		\label{bi-uncompute-near-term}
	\end{subfigure} 
	\caption{Gates used in this work.  $\ket{A}$ is an ancillae in the state $\frac{1}{\sqrt{2}}(\ket{0} + e^{\frac{i \cdot \pi}{4}}\ket{1})$. }
	\label{QCLA-logic-gates}
\end{figure}

To overcome shortcomings in existing works, this work proposes a novel QCLA for NISQ and fault-tolerant machines.  In-place QCLA and out-of-place QCLA implementations for the design are shown.  Both designs enjoy reduced T gate cost and qubit cost compared to the existing works.  The proposed QCLAs use a proposed uncomputation gate with a T-count of 3.   The proposed uncomputation gate allows the QCLAS to be used on NISQ machines.  The proposed QCLA designs are based on the NOT gate, CNOT gate, Toffoli gate, logical AND gate and the novel uncomputation gate.  The logical AND gate depicted in Figure \ref{QCLA-logic-gates} is presented in \cite{Gidney20184TgateToffolibuild}.
The existing designs in \cite{Draper2006qcla} are based solely on CNOT, NOT and Toffoli gates. 

The proposed work modifies the design methodologies in \cite{Draper2006qcla} by replacing Toffoli gate with logical-AND gate and uncomputation gate pairs into the design where possible. The proposed uncomputation gate is used to avoid the measurement operation required by the uncomputation gate proposed in \cite{Gidney20184TgateToffolibuild}. We seek to avoid the measurement operation because (i) the errors associated with the operation (such as SPAM errors) and (ii) the increased length of time to perform measurement compared to gates on NISQ machines. In consequence, the number of T gates thereby the total number of quantum gates used in the proposed design is significantly reduced compared to existing work.

Further, quantum computing has been applied to image orientation problems and image pattern recognition  \cite{Beach2004motivation} \cite{Venegas-AndracaS2016morivation}. To implement quantum image processing algorithms, images must be encoded on quantum hardware and circuits to manipulate image representations must be designed. In this work, we illustrate an example of a quantum circuit for quantum image processing by presenting quantum circuits for bilinear interpolation based on the proposed QCLAs.

\begin{figure}[htbp]
	\centering
	\small
	\[
	\Qcircuit @C=0.7em @R=0.5em @!R{
		\lstick{\ket{0}} & \qw & \qw  & \qw & \qw & \qw & \qw & \qw & \qw & \targ & \targ & \qw & \rstick{\ket{s_0}}\\
		\lstick{\ket{a_0}} & \ctrl{2} & \qw & \qw & \qw & \qw & \qw & \qw & \qw & \qw & \ctrl{-1} & \qw & \rstick{\ket{a_0}}\\
		\lstick{\ket{b_0}} & \ctrl{1} & \qw & \qw & \qw & \qw & \qw & \qw & \qw & \ctrl{-2} & \qw & \qw & \rstick{\ket{b_0}} \\
		\lstick{\ket{A}}   &  & \qw & \qw  & \ctrl{3}  & \qw & \qw & \qw & \qw & \targ & \qw & \qw & \rstick{\ket{s_1}}\\
		\lstick{\ket{a_1}} & \ctrl{2} & \ctrl{1} & \qw & \qw & \qw & \qw & \qw & \qw & \qw & \ctrl{1} & \qw & \rstick{\ket{a_1}}\\ 
		\lstick{\ket{b_1}} & \ctrl{1} & \targ & \qw & \ctrl{1}& \qw & \qw & \qw &\qw & \ctrl{-2} & \targ & \qw & \rstick{\ket{b_1}}\\ 
		\lstick{\ket{A}}   & & \qw & \qw & \targ & \ctrl{4} & \ctrl{3} & \qw & \qw & \targ & \qw & \qw & \rstick{\ket{s_2}}\\  
		\lstick{\ket{a_2}} & \ctrl{2} & \ctrl{1}   & \qw & \qw & \qw & \qw & \qw & \qw & \qw & \ctrl{1} & \qw & \rstick{\ket{a_2}}\\
		\lstick{\ket{b_2}} & \ctrl{2} & \targ & \ctrl{2} & \qw & \qw & \ctrl{2} & \qw & \ctrl{2} & \ctrl{-2} & \targ & \qw & \rstick{\ket{s_2}}\\
		\lstick{\ket{A}}   &  &  &  & \qw & \ctrl{4}  & \qw &\qw & \qw &  & & & &\\
		\lstick{\ket{A}}   &  & \qw & \qw  & \ctrl{3}  & \qw & \targ &\qw & \qw & \targ & \qw & \qw & \rstick{\ket{s_3}}\\
		\lstick{\ket{a_3}} & \ctrl{2} & \ctrl{1} &\qw  & \qw & \qw & \qw &\qw & \qw & \qw & \ctrl{1} & \qw & \rstick{\ket{a_3}}\\
		\lstick{\ket{b_3}} & \ctrl{1} & \targ & \ctrl{-2} & \ctrl{1} & \qw & \qw & \qw & \ctrl{-2} & \ctrl{-2} &\targ & \qw & \rstick{\ket{b_3}}\\ 
		\lstick{\ket{A}}   &  & \qw & \qw  & \targ & \targ &\qw & \qw & \qw & \qw & \qw & \qw & \rstick{\ket{s_4}}\\
	}
	\]
	\caption{Proposed out-of-place QCLA for the case of adding two $4$ bit values $a$ and $b$.}
	\label{OOP-NearTerm}
\end{figure}
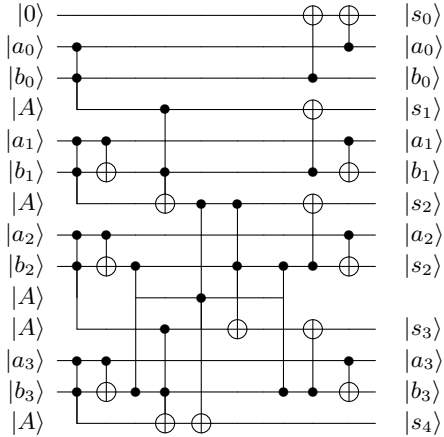

\section{Proposed Design of Out-of-Place QCLA Circuit} 

The $8$ step methodology to implement proposed out-of-place QCLA is generic and can be used to implement a QCLA circuit of any size.  The methodology is a modified version of the design methodology presented in \cite{Draper2006qcla}.  An example of the proposed QCLA is shown in Figure \ref{OOP-NearTerm}.  The proposed QCLA saves resources by using the temporary logical-AND gate and proposed uncomputation gate where possible.   

The proposed out-of-place QCLA circuit operates as follows: Given 2 values $a$ and $b$ (each $n$ bits wide) stored in quantum registers $A$ and $B$ as well as $n+1$ ancillae stored in register $X$.  $X_0$ is set to $0$ and the remaining locations are set to $A$ $\left( \text{ where } A = \frac{1}{\sqrt{2}}\left(\ket{0} + e^{\frac{i \pi}{4}}\ket{1}\right)\right)$. The QCLA require a register $Z$ with $n-w(n)-\lfloor log(n) \rfloor$ elements (where $w (n) = n - \sum_{y=1}^{\infty} \lfloor \frac{n}{2^y} \rfloor$ and is the number of ones in the binary expansion of $n$) all set to $A$ $\left( \text{ where } A = \frac{1}{\sqrt{2}}\left(\ket{0} + e^{\frac{i \pi}{4}}\ket{1}\right)\right)$.  At the end of computation, $A$ and $B$ are restored to their initial values and $X$ will contain the sum of the addition of $a$ and $b$.  The QCLA will restore $Z$ to $\frac{1}{\sqrt{2}}\left(\ket{0} + e^{\frac{i \pi}{4}}\ket{1}\right)$.  These qubits can be set to computational basis values for reuse as ancillae.  The steps to implement QCLA are shown along with an illustrative example of the QCLA circuit in Figure \ref{OOP-NearTerm}

\section{Proposed Design of In-Place QCLA Circuit} 

\begin{figure}[htbp]
	\centering
	\small
	\[
	\Qcircuit @C=0.2em @R=0.5em @!R{
		\lstick{\ket{a_0}} & \ctrl{2} & \ctrl{1} & \qw & \qw & \qw & \qw & \qw & \qw & \qw & \qw & \qw & \qw & \qw & \qw & \qw &\qw & \qw & \qw & \qw  &  \qw &  \qw & \ctrl{2} & \qw & \qw & \rstick{\ket{a_0}} \\
		\lstick{\ket{b_0}} & \ctrl{1} & \targ & \qw & \qw & \qw & \qw & \qw & \qw & \qw &\qw & \qw & \qw & \qw & \qw & \targ & \qw  & \qw & \qw & \qw & \qw & \qw &  \ctrl{1} & \targ & \qw & \rstick{\ket{s_0}} \\
		\lstick{\ket{A}}   &  & \qw & \qw & \qw & \qw & \qw & \ctrl{3} & \qw &  \qw &\qw & \qw & \qw & \qw & \ctrl{2} & \qw & \qw & \qw & \qw & \ctrl{2} & \qw &  \qw & \qw &  &  &\\
		\lstick{\ket{a_1}} & \ctrl{2} & \ctrl{1} & \qw & \qw & \qw & \qw & \qw &  \qw &  \qw & \qw & \qw & \qw & \qw & \qw & \qw & \ctrl{1} & \qw & \qw & \qw & \qw &  \ctrl{1} & \ctrl{2} & \qw & \qw & \rstick{\ket{a_1}}\\
		\lstick{\ket{b_1}} & \ctrl{1} & \targ & \qw & \qw & \qw & \qw & \ctrl{1} &  \qw & \qw & \qw & \qw & \qw & \qw & \targ & \targ & \targ &\qw & \qw &\ctrl{1} & \qw & \targ &  \ctrl{1} & \targ & \qw & \rstick{\ket{s_1}}\\
		\lstick{\ket{A}}   &  & \qw & \qw & \qw & \qw & \qw & \targ& \ctrl{3} & \qw & \qw & \ctrl{3} & \qw & \qw & \ctrl{2} & \qw & \qw & \qw & \ctrl{2} & \targ & \qw &  \qw & \qw &  &  &\\
		\lstick{\ket{a_2}} & \ctrl{2} & \ctrl{1} & \qw & \qw & \qw & \qw & \qw &  \qw & \qw & \qw & \qw & \qw & \qw & \qw & \qw & \ctrl{1} & \qw & \qw & \qw & \qw &  \ctrl{1} & \ctrl{2} & \qw & \qw & \rstick{\ket{a_2}} \\
		\lstick{\ket{b_2}} & \ctrl{2} & \targ & \ctrl{2} & \qw & \qw & \qw & \qw &  \qw & \qw & \qw & \ctrl{2} & \qw &\ctrl{2} &\targ & \targ & \targ & \qw & \ctrl{2} & \qw & \qw & \targ & \ctrl{2} & \targ & \qw & \rstick{\ket{s_2}}\\
		\lstick{\ket{A}}   &  &  &  & \qw & \qw & \qw & \qw  & \ctrl{4} &  \qw & \qw & \qw &\qw & \qw &  &  &  &  &  &  &  &  &  &  &\\
		\lstick{\ket{A}}   &  & \qw & \qw & \qw & \qw & \qw & \ctrl{2}  & \qw &  \qw & \qw & \targ & \qw & \qw & \ctrl{2} &\qw & \qw & \qw & \targ & \qw & \qw &  \qw & \qw &  &  &\\
		\lstick{\ket{a_3}} & \ctrl{2} & \ctrl{1} & \qw & \qw & \qw & \qw & \qw & \qw &  \qw & \qw & \qw & \qw & \qw & \qw & \qw & \qw & \qw & \qw  & \qw & \qw &  \qw & \qw & \qw &\qw &  \rstick{\ket{a_3}}\\
		\lstick{\ket{b_3}} & \ctrl{1} & \targ & \ctrl{-2} & \qw & \qw & \qw & \ctrl{1} & \qw &  \qw & \qw & \qw & \qw & \ctrl{-2} & \targ & \qw & \qw & \qw & \qw & \qw & \qw &\qw & \qw &  \qw & \qw  & \rstick{\ket{s_3}}\\
		\lstick{\ket{A}}   &  & \qw & \qw &\qw & \qw & \qw &\targ &\targ &\qw & \qw & \qw & \qw & \qw & \qw & \qw & \qw & \qw & \qw & \qw & \qw & \qw  & \qw & \qw &\qw & \rstick{\ket{s_4}} \\
	}
	\]
	\caption{In-place QCLA for the case of adding two $4$ bit values $a$ and $b$.}
	\label{IP-NearTerm}
\end{figure}
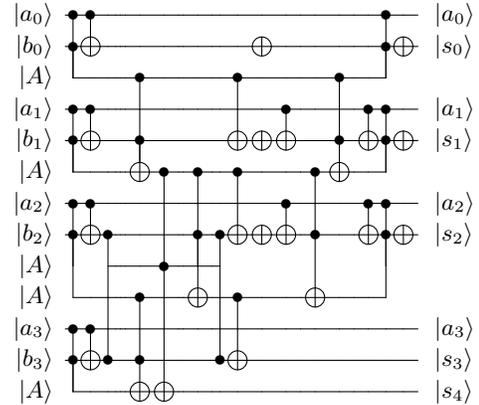

The $16$ step methodology to implement proposed in-place QCLA is generic and can be used to implement a QCLA circuit of any size.  The methodology is a modified version of the design methodology presented in \cite{Draper2006qcla}.  An example of the proposed QCLA is shown in Figure \ref{IP-NearTerm}.  The proposed QCLA saves resources by using the temporary logical-AND gate and proposed uncomputation gate where possible. 

The proposed QCLA operates as follows: Given 2 values $a$ and $b$ (each $n$ bits wide) stored in quantum registers $A$, $B$ and an $n$ bit register $Z$ of ancillae set to $\frac{1}{\sqrt{2}}\left(\ket{0} + e^{\frac{i \pi}{4}}\ket{1}\right)$.  Lastly, QCLA requires a $n-w(n)-\lfloor log(n) \rfloor$ (where $w(n) = n - \sum_{y=1}^{\infty} \bigl\lfloor \frac{n}{2^y} \bigr\rfloor$ and is the number of ones in the binary expansion of $n$) bit register of ancillae ($X$) set to $\frac{1}{\sqrt{2}}\left(\ket{0} + e^{\frac{i \pi}{4}}\ket{1}\right)$.  At the end of computation, $A$ is restored to their initial values and $B$ will contain sum bits $0$ through $n-1$ of the addition of $a$ and $b$.  At the end of computation, $Z[n]$ will contain the sum bit $s_n$.  The QCLA will restore the remaining locations in $Z$ to $\frac{1}{\sqrt{2}}\left(\ket{0} + e^{\frac{i \pi}{4}}\ket{1}\right)$.  The QCLA will restore $X$ to $\frac{1}{\sqrt{2}}\left(\ket{0} + e^{\frac{i \pi}{4}}\ket{1}\right)$.

\section{Performance of Proposed QCLA circuits}

\begin{table}[tbhp]
	\centering
	\small
	\caption{Cost Comparison of Out-of-place QCLAs}
	\begin{tabular}{lcc}
		\\ \midrule												
		Design	&	& T-count Equation	\\		\cmidrule{1-1} \cmidrule{3-3}			
		Draper et al. (\cite{Draper2006qcla})	&	& $35n-21w(n)-21\lfloor log(n)\rfloor-7$      	\\
		Thapliyal et al. (\cite{Thapliyal2013qcla} )		&   &   $35n-14$        \\
		Babu et. al. (\cite{Babu2013QCLAoutofplace})* & & $54 \cdot n$ \\
		Proposed	&   & $25n-14w(n)-14\lfloor log(n)\rfloor-7$  
		\\	\bottomrule	
		\multicolumn{3}{l}{$w(n) = n - \sum_{y=1}^{\infty} \bigl\lfloor \frac{n}{2^y} \bigr\rfloor$} \\
		\multicolumn{3}{p{3in}}{* Circuits modified to remove garbage output.  We use the methodology in \cite{Bennett1973trashremoval} to remove the garbage output.}\\
	\end{tabular}
	\label{Equations for Out-of-place T-count}
\end{table}

\begin{table}[tbhp]
	
	\centering
	\small
	\caption{Cost Comparison of In-Place QCLAs}
	\begin{tabular}{p{.6in}cp{2.25in}}
		\\ \midrule												
		Design	&	& T-count Equation	\\		\cmidrule{1-1} \cmidrule{3-3}			
		Draper et al. (\cite{Draper2006qcla})	&	&	$70n-21w(n)-21\lfloor log(n)\rfloor-21w(n-1)-21\lfloor log (n-1) \rfloor-49$              \\ 
		Thapliyal et al. (\cite{Thapliyal2013qcla})	&   &   $\frac{203}{4}n-28$   \\
			Cheng et al. (\cite{Cheng2002QCLAinplace}) & & $\frac{14}{6} n^3 + \frac{21}{6} n^2 - \frac{49}{6} n$ \\
		Proposed &   & $46n - 14w(n) -  14\lfloor log(n) \rfloor - 14w(n-1) -  14\lfloor log(n-1) \rfloor - 36$  \\	\bottomrule	
		\multicolumn{3}{l}{$w(n) = n - \sum_{y=1}^{\infty} \bigl\lfloor \frac{n}{2^y} \bigr\rfloor$} \\
	\end{tabular}
	\label{Equations for In-place T-count}
\end{table}

Table \ref{Equations for Out-of-place T-count} indicates the proposed out-of-place QCLA has T-count cost of order $\mathcal{O}(n)$.  The proposed QCLA requires $53.70 \%$ fewer T gates than the design by Babu et al., $28.57 \%$ fewer T gates than the designs by Draper et al. and Thapliyal et al. 

Table \ref{Equations for In-place T-count} indicates that the proposed in-place QCLA has a T-count cost of order $\mathcal{O}(n)$.  The proposed in-place QCLA requires $34.29 \%$ fewer T gates than the designs by Draper et al., $9.36 \%$ fewer T gates than the designs by Thapliyal et al. and has a polynomial factor improvement over the work in Cheng et al.

\section{Application in Quantum Image Processing}
\label{IT-Pro-BI}

\begin{figure}
\centering

\begin{subfigure}[tbhp]{2.25in}
\includegraphics[width = 2.25in]{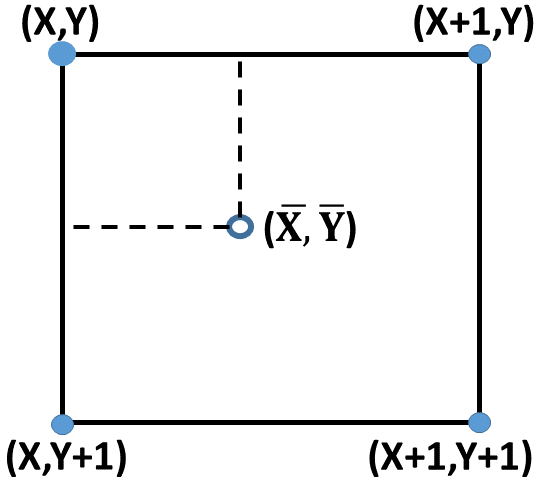}
\caption{Coordinate mapping of a bilinear interpolated image. }
\label{BI-linear-int}
\end{subfigure}
\\
\begin{subfigure}[tbhp]{3in}
\flushright
\scriptsize 
\[
\Qcircuit @C=0.2em @R=0.5em @!R{
\lstick{\ket{10 \cdots 01}} & \qw &\ctrl{1} &\ctrl{2} & \qw & \qw & \qw & \qw & \qw & \qw & \qw & \qw & \qw & \qw & \qw & \qw & \qw & \rstick{\ket{10 \cdots 01}}\\
\lstick{\ket{Y_{n-1:0} }} & \ctrl{4} &\gate{-} & \qw &\ctrl{3} &\qw &\ctrl{5}  & \qw & \qw & \qw & \qw & \qw & \qw & \qw & \qw & \qw & \qw& \rstick{\ket{Garbage} }\\
\lstick{\ket{X_{n-1:0} }} & \qw & \ctrl{4} &\gate{-} &\ctrl{5} & \ctrl{3} &\qw & \qw & \qw & \qw & \qw & \qw & \qw & \qw & \qw & \qw & \qw& \rstick{\ket{Garbage} }\\
\lstick{\ket{Y_{m-1:n} }} & \qw & \qw & \qw & \qw &\qw &\qw &\qw & \qw & \qw & \qw & \qw & \qw & \qw & \qw & \qw & \qw & \rstick{\ket{\overline{Y}}}\\
\lstick{\ket{X_{m-1:n} }} & \qw & \qw & \qw & \qw & \qw &\qw &\qw & \qw & \qw & \qw & \qw & \qw & \qw & \qw & \qw & \qw& \rstick{\ket{\overline{X}}}\\
\lstick{\ket{0}} & \targ & \qw & \qw & \qw & \ctrl{3} & \qw & \ctrl{3} & \qw & \qw & \qw & \qw & \qw & \qw & \qw & \qw & \qw & \rstick{\ket{Garbage} }\\
\lstick{\ket{0}} & \qw & \targ & \qw & \qw & \qw & \ctrl{3} & \ctrl{4} & \qw & \qw & \qw & \qw & \qw & \qw & \qw & \qw & \qw& \rstick{\ket{Garbage} }\\
\lstick{\ket{0}} & \qw & \qw & \qw &\gate{x} & \qw & \qw & \qw & \qw & \qw & \qw & \qw &\ctrl{4} & \qw & \qw & \qw & \qw &\rstick{\ket{Garbage} }\\
\lstick{\ket{0}} & \qw & \qw & \qw &\qw &\gate{x} & \qw & \qw & \qw & \qw & \qw &\ctrl{4} & \qw & \qw & \qw & \qw & \qw &\rstick{\ket{Garbage} }\\
\lstick{\ket{0}} & \qw & \qw & \qw & \qw & \qw &\gate{x} & \qw & \qw & \qw &\ctrl{4} & \qw & \qw & \qw & \qw & \qw & \qw &\rstick{\ket{Garbage} }\\
\lstick{\ket{0}} & \qw & \qw & \qw & \qw & \qw & \qw &\gate{x} & \qw &\ctrl{4} & \qw & \qw & \qw & \qw & \qw & \qw & \qw &\rstick{\ket{Garbage} }\\
\lstick{\ket{C_{Y,X}}} & \qw & \qw & \qw & \qw & \qw & \qw & \qw & \qw & \qw & \qw & \qw &\ctrl{7} & \qw & \qw & \qw & \qw& \rstick{\ket{C_{Y,X}}}\\
\lstick{\ket{C_{Y+1,X}}} & \qw & \qw & \qw & \qw & \qw & \qw & \qw & \qw & \qw & \qw &\ctrl{5} & \qw & \qw & \qw & \qw  & \qw& \rstick{\ket{C_{Y+1,X}}}\\
\lstick{\ket{C_{Y,X+1}}} & \qw & \qw & \qw & \qw & \qw & \qw & \qw & \qw & \qw & \ctrl{3} & \qw & \qw & \qw & \qw & \qw  & \qw& \rstick{\ket{C_{Y,X+1}}}\\
\lstick{\ket{C_{Y+1,X+1}}} & \qw & \qw & \qw & \qw & \qw & \qw & \qw & \qw & \ctrl{1} & \qw  & \qw & \qw & \qw & \qw & \qw & \qw& \rstick{\ket{C_{Y+1,X+1}}}\\
\lstick{\ket{0}} & \qw & \qw & \qw & \qw & \qw & \qw & \qw & \qw & \gate{x} & \qw & \qw & \qw & \qw & \qw &\ctrl{3} & \qw& \rstick{\ket{Garbage} }\\
\lstick{\ket{0}} & \qw & \qw & \qw & \qw & \qw & \qw & \qw & \qw & \qw &\gate{x} & \qw & \qw & \qw &\ctrl{2} & \qw & \qw &\rstick{\ket{Garbage} }\\
\lstick{\ket{0}} & \qw & \qw & \qw & \qw & \qw & \qw & \qw & \qw & \qw &\qw & \gate{x} & \qw &\ctrl{1} & \qw & \qw & \qw &\rstick{\ket{Garbage} }\\
\lstick{\ket{0}} & \qw & \qw & \qw & \qw & \qw & \qw & \qw & \qw & \qw & \qw &\qw & \multigate{1}{x} & \multigate{1}{+} & \multigate{1}{+} & \multigate{1}{+} & \qw &\rstick{\ket{C_{\overline{Y},\overline{X}}}}\\
\lstick{\ket{0}} & \qw & \qw & \qw & \qw & \qw & \qw & \qw & \qw & \qw & \qw &\qw & \ghost{x} & \ghost{+} & \ghost{+} & \ghost{+} & \qw &\rstick{\ket{Garbage} }\\
}
\]
\caption{Quantum bilinear interpolation circuit for scaling down by a value $n$. }
\label{bilinear-scale-down-image-complete}
\end{subfigure}
\caption{Application of quantum arithmetic circuits in image processing.}
\label{application-IT-Pro}
\end{figure}

We will now illustrate how the proposed QCLAs can be used to implement a quantum circuit for bilinear interpolation.  Interpolation is of interest because it is used in image processing operations such as zooming, rotations, and resampling \cite{Hakran2011whyBI} \cite{YinChen2009whyBI}.  Bilinear interpolation is a well established method for scaling images \cite{Gonzalez1992DIPbook}.  Bilinear interpolation uses linear interpolation to sequentially perform interpolation for each pixel location variable $x$ and $y$.  Figure \ref{BI-linear-int} shows an example coordinate mapping of a bilinear interpolated pixel at original location $(X, Y)$.  Further details about bilinear interpolation can be found in \cite{Gonzalez1992DIPbook}.

\begin{figure}
	\centering
	\includegraphics[width=0.8\linewidth]{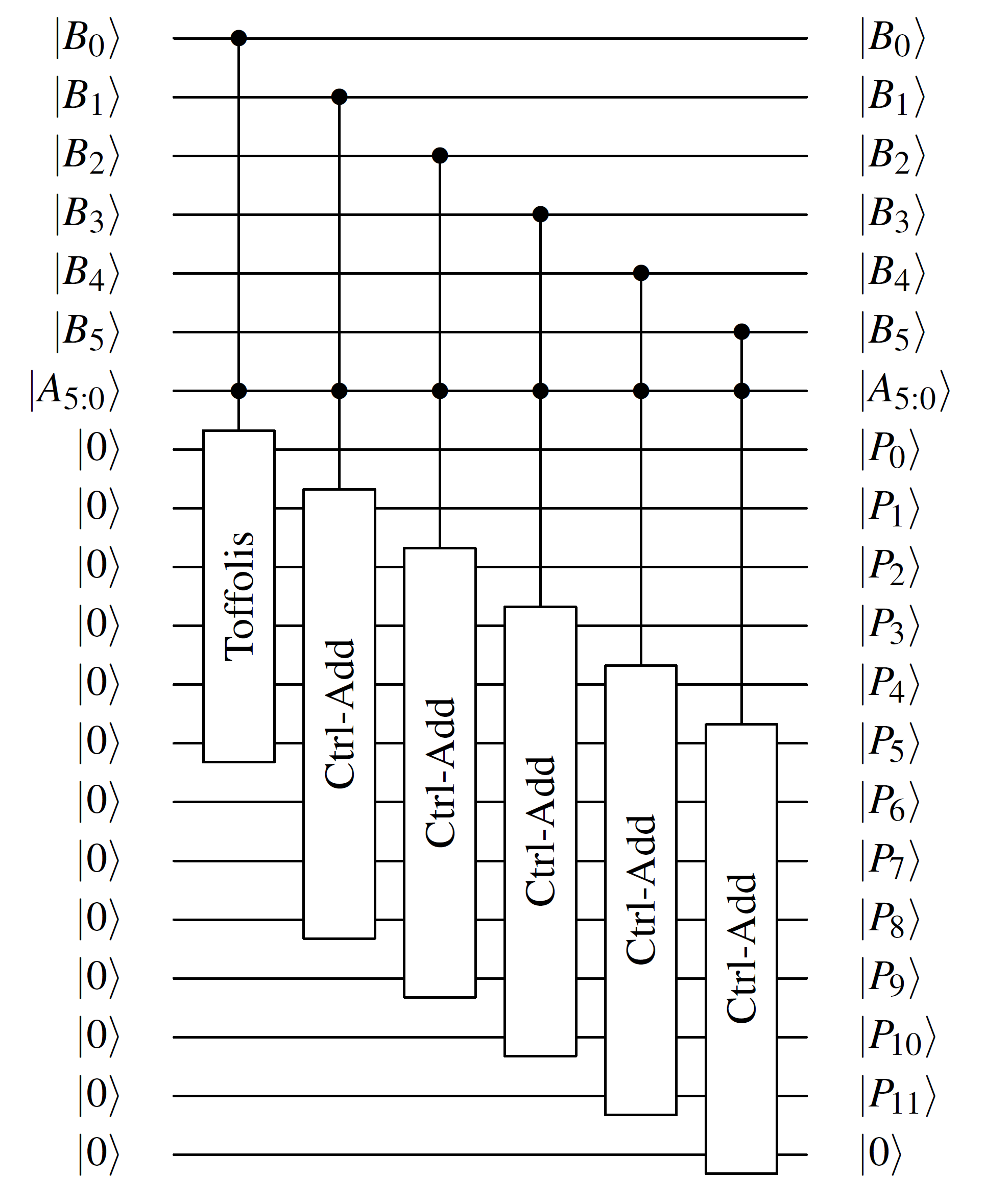}
\caption{Example of the proposed quantum integer multiplication circuit for the multiplication of two $6$-bit inputs $\ket{A}$ and $\ket{B}$.  $\ket{P}$ is the product of the values in $\ket{A}$ and $\ket{B}$.  }
\label{ISVLSI-INT-multiply-pic}
\end{figure}

The QCLA circuits can be used to implement quantum circuits for bilinear interpolation.  We can build quantum circuits for scaling down and for scaling up an image by an integer $n$.  Scaling up by $n$ will increase an image by $2^n$ and scaling down by $n$ will decrease an image by $2^n$.  We show the complete quantum bilinear interpolation circuit for the scale down operation in Figure \ref{bilinear-scale-down-image-complete}.  Both quantum bilinear interpolation use a (i) quantum adder, (ii) quantum subtractor and (iii) quantum multiplier. The required block of quantum adder, quantum subtractor and quantum multiplier can be easily designed from the proposed quantum carry lookahead adders \cite{Edgard2018bilinear}. As an illustrative example, we have shown a quantum integer multiplication circuit that is optimized for T-count and qubits \cite{Edgard2019multiplier}. The building blocks of the quantum integer multiplication circuit are (i)  quantum \textit{Ctrl-Add} circuit  and (ii) arrays of Toffoli gates. The quantum \textit{Ctrl-Add} circuit is based on a resource efficient quantum carry look-ahead adders presented in this work.  Figure \ref{ISVLSI-INT-multiply-pic} shows an example of the proposed multiplier for the case of multiplying two $6$ bit integers.  In Figure \ref{ISVLSI-INT-multiply-pic}, quantum registers $\ket{A}$ and $\ket{B}$ contain the inputs $a$ and $b$ to be multiplied.  The proposed quantum integer multiplication circuit implements the shift and add multiplication algorithm.  The placement of the \textit{Ctrl-Add} circuits eliminates the need for gates to implement the shifting operation (see Figure \ref{ISVLSI-INT-multiply-pic}).  To reduce gate cost \textit{Ctrl-Add} circuits are replaced by Toffoli gate arrays where possible.  
         
We follow the procedures in \cite{Edgard2018bilinear} to implement both quantum bilinear interpolation circuits. Both circuits take the original pixel's position $\ket{Y}, \ket{X}$ and color information $\ket{C_{Y,X}}$ as inputs.  The color information for pixels at locations ($y+1,x$), ($y,x+1$) and ($y+1,x+1$) are also inputs.  The circuitry outputs the scaled pixel's location $\ket{\overline{Y}}$ and $\ket{\overline{X}}$ and its color information $\ket{C_{\overline{Y},\overline{X}}}$.  
  
 We determine the location information for the scaled pixel without gates by either appending ancillae or truncating the input location values.  The quantum bilinear interpolation circuit to scale down an image evaluates expression \ref{blabs:1} for the scaled color value:

{\small 
\begin{equation}
\left[ \begin{matrix}
(2^n - \ket{\widetilde{Y}}) \cdot (2^n - \ket{\widetilde{X}}) \cdot C_{Y,X} + \\
\ket{\widetilde{Y}} \cdot (2^n - \ket{\widetilde{X}}) \cdot C_{Y+1,X} + \\
(2^n - \ket{\widetilde{Y}} \cdot \ket{\widetilde{X}} ) \cdot C_{Y,X+1} + \\
\ket{\widetilde{Y}} \cdot \ket{\widetilde{X}} ) \cdot C_{Y+1,X+1} \\
\end{matrix} \right] \div 2^{2 \cdot n}
\label{blabs:1}
\end{equation}
}

Where $ \widetilde{Y} = Y_{n-1:0}$ and $ \widetilde{X} = X_{n-1:0}$.  The quantum bilinear interpolation circuit to scale up an image evaluates expression \ref{blabs:2} for the scaled color value: 

{\small 
\begin{equation}
\left[ \begin{matrix}
\left(2^m - \ket{\widetilde{Y}} \right) \cdot \left(2^m - \ket{\widetilde{X}} \right) \cdot C_{Y,X} + \\
\left(\ket{\widetilde{Y}} \right) \cdot (2^m -\left( \ket{\widetilde{X}} \right) \cdot C_{Y+1,X} + \\
\left(2^m - \ket{\widetilde{Y}} \right) \cdot \left( \ket{\widetilde{X}} \right) \cdot C_{Y,X+1} + \\
\left( \ket{\widetilde{Y}} \right) \cdot \left( \ket{\widetilde{X}} \right) \cdot C_{Y+1,X+1} \\
\end{matrix} \right] \div 2^{2 \cdot m}
\label{blabs:2}
\end{equation}
}

Where $ \widetilde{Y} = \overline{Y_{m+n-1:n-1}}$ and $ \widetilde{X} = \overline{X_{m+n-1:n-1}}$.

Details of the design of the quantum bilinear interpolation circuits for the scale down operation and for the scale up operation are illustrated in \cite{Edgard2018bilinear}.  We determine $\left( C_{\overline{Y},\overline{X}} \right) $ for both circuits without division by truncating the quantum register containing the result of computation (see Figure \ref{bilinear-scale-down-image-complete}).

\section{Conclusion}

In this work, we propose quantum circuits for carry lookahead addition for NISQ machines.  We propose designs for an in-place QCLA and out-of-place QCLA.  The proposed QCLAs are optimized for low T gate cost.  The proposed QCLAs are based on the NOT gate, the CNOT gate, the Toffoli gate, the logical AND gate and the proposed uncomputation gate.  These designs are compared and shown to have reduced T gate and therefore total number of quantum gates compared to the existing work.  The proposed QCLAs also enjoy qubit cost savings compared to existing work.  We conclude that the proposed in-place QCLA and out-of-place QCLA can be used in larger quantum data-path circuits in NISQ machines or when fault-tolerant quantum circuit design is not possible.  We also illustrate the application of the proposed QCLAs in image processing through the example of circuits for quantum  bilinear interpolation.

\bibliographystyle{IEEEtran}
\bibliography{references.bib}

\end{document}